\documentclass[aoas,preprint,graybox,envcountchap,sectrefs]{imsart}

\RequirePackage[OT1]{fontenc}
\RequirePackage{amsthm,amsmath}
\RequirePackage[numbers]{natbib}
\RequirePackage[colorlinks,citecolor=blue,urlcolor=blue]{hyperref}
\RequirePackage{hypernat}

\usepackage{graphicx}
\usepackage{parskip}

\usepackage{amsfonts}
\usepackage{natbib}
\usepackage{algorithm}
\usepackage{algorithmic}
\usepackage{subfigure}

\usepackage{latexsym}

\newtheorem{theorem}{Theorem}[section]

\newtheorem{lemma}{Lemma}[section]
\newtheorem{corollary}{Corollary}[section]
\newtheorem{remark}{Remark}[section]
\newtheorem{example}{Example}[section]

\newcommand{\R}{\mathbb{R}}

\newcommand{\PP} {{  \rm I\hskip-0.22em P}}

\newcommand{\EE} {{\rm I\hskip-0.48em E}}

\usepackage{natbib}

\arxiv{math.PR/0000000}

\startlocaldefs
\numberwithin{equation}{section}
\theoremstyle{plain}
\endlocaldefs

\begin{document}

\begin{frontmatter}
\title{The Lasso, correlated design, and improved oracle inequalities\protect\thanksref{T1}}
\runtitle{Lasso with correlated design}
\thankstext{T1}{Research supported by SNF 20PA21-120050}

\begin{aug}
\author{\fnms{Sara} \snm{van de Geer}\ead[label=e1]{geer@stat.math.ethz.ch}}
\and
\author{\fnms{Johannes} \snm{Lederer}\ead[label=e2]{lederer@stat.math.ethz.ch}}

%
%\thankstext{t1}{Some comment}
%\thankstext{t2}{First supporter of the project}
%\thankstext{t3}{Second supporter of the project}
\runauthor{S. van de Geer et al.}

\affiliation{ETH Z\"urich}

\address{Seminar for Statistics\\
ETH Z\"urich\\
R\"amistrasse 101\\
8092 Z\"urich\\
\printead{e1}\\
\phantom{E-mail:\ }\printead*{e2}}

\end{aug}

\begin{abstract}
We study high-dimensional linear models
and the $\ell_1$-penalized least squares estimator, also
known as the Lasso estimator. In literature, oracle inequalities have been derived
under restricted eigenvalue or compatibility conditions. In this paper, we complement
this with entropy conditions which allow one to improve the dual
norm bound, and demonstrate how this leads to new oracle inequalities.
The new oracle inequalities show that a smaller choice for the tuning parameter 
and a trade-off between $\ell_1$-norms and small
compatibility constants are possible. This implies, in particular for
correlated design,  improved bounds for the prediction error
of the Lasso estimator as compared to the methods based on restricted eigenvalue or
compatibility conditions only.\end{abstract}



\end{frontmatter}

\section{Introduction}\label{intro.section}
We derive oracle inequalities for the Lasso estimator for various designs.
Results in literature are generally based on
restricted eigenvalue
 or compatibility conditions (see Section \ref{definitions.section} for definitions).
 We refer to
 \cite{bickel2009sal}, \cite{Bunea:06},  \cite{Bunea:07a}, \cite{Bunea:07b}, 
  \cite {koltch09a}, \cite {vandeG07},  \cite {vandeG08}.
See also  \cite{ BvdG2011} and the references therein.
 In a sense, compatibility or restricted eigenvalue conditions and
 the so-called dual norm bound we describe below belong together. In contrast, if compatibility constants or restricted eigenvalues
 are very small, the design may have high correlations, and then the dual norm bound is too rough. In this paper, we discuss an approach that joins both
 situations. 
 The work is a follow-up of \cite{vdG2007}. It combines results of the latter with the
parallel developments in the area based on the dual norm bound.

We consider an input space ${\cal X}$ and $p$ feature mappings
$ \psi_j : {\cal X} \rightarrow \R$, $j=1 , \ldots , p $.
We let $( x_1 , \ldots , x_n )^T \in {\cal X}^n $ be a given  input vector,
and ${\bf Y} := ( {\bf Y}_1 , \ldots , {\bf Y}_n )^T \in \R^n $ be an output vector, and
consider the linear model
$${\bf Y} = \sum_{j=1}^p \psi_j \beta_j^0 + \epsilon, $$
with $\epsilon \in \R^n$ a noise vector, and $\beta^0 \in \R^p$
a vector of unknown coefficients. Here, with some abuse of notation,
$\psi_j$ denotes the vector
$\psi_j = ( \psi_j (x_1) , \ldots , \psi_j (x_n))^T$. The design matrix is
${\bf X} := ( \psi_1 , \ldots , \psi_p )$ and the Gram matrix is
$$\hat  \Sigma :=  {\bf X}^T {\bf X} / n .$$ Throughout, we assume
that
$\sum_{i=1}^n \psi_j^2 (x_i) \le n $ for all $j$.
 
We write a linear function with
coefficients $\beta$ as $f_{\beta} := \sum_{j=1}^p  \psi_j\beta_j$, $\beta \in \R^p$.
 The Lasso estimator is
 $$\hat \beta := \arg \min_{\beta} \biggl \{
 \| {\bf Y } - f_{\beta} \|_2^2 / n +  \lambda \| \beta \|_1  \biggr \}.
 $$
We denote the estimator of the regression function
$f^0:= f_{\beta^0}$ by $  \hat f : = f_{\hat \beta}  $.
 
 Oracle results using compatibility or restricted eigenvalue conditions are based on the
 dual norm bound
 $$\sup_{\| \beta  \|_1 = 1 } |\epsilon^Tf_{\beta}  | /n=
\max_{1 \le j \le p } |  \epsilon^T \psi_j | / n  . $$
Let us define
$$\| f_{\beta}  \|_n^2 := \sum_{i=1}^n 
 f_{\beta}^2 (x_i)  / n = \beta^T
 \hat \Sigma  \beta . $$
The point we make in this paper is that the dual norm bound does not take into account
 possible small values for $\| f_{\hat \beta} - f_{\beta^0} \|_n$. 
 Our results are based on bounds for
 $$\sup_{\| \beta  \|_1 \le 1 , \ \| f_{\beta} 
  \|_n \le R } |\epsilon^T f_{\beta}  | /n  $$
 as function of $R>0$. We then apply these to $\hat \beta - \beta^0$ (or
 $\beta^0$ here replaced by a sparse approximation). 
 We use an
 improvement of the dual norm bound, and
 show in Theorem \ref{main.theorem} the consequences.
 The main observation here is that with highly
 correlated design, one can generally take the tuning parameter $\lambda$ of
 much smaller order than the usual $\sqrt {\log p / n }$. 
 Moreover, small compatibility constants may be traded off against
 the $\ell_1$-norm of  the coefficients of an oracle.
  
 \section{Organization of the paper}
 In Section \ref{definitions.section}, we present our notation, and the definitions
 of compatibility constants and restricted eigenvalues.
 Section \ref{main.section} contains the main result,
 based on a pre-assumed
 improvement of the dual norm bound. In Section \ref{dudley.section}, we present a result from
 empirical process theory, which shows that the improvement
 of the dual norm bound used in Section \ref{main.section} holds
 under entropy conditions on ${\cal F}:= \{ f_{\beta} : \ \| \beta \|_1 = 1 \}$.
 In Section \ref{measurecorr.section}, we first
 give a geometrical interpretation of the compatibility constant
 and discuss the relation with eigenvalues. 
  The next question to address is then how to read off  the entropy conditions
 directly from the design.
 We show that a 
 Gram matrix with strongly decreasing eigenvalues leads to a small
 entropy of ${\cal F}$.
Alternatively, we derive an 
 an entropy bound for ${\cal F} $ based on the covering number
 of the design $\{ \psi_j \}$, a result much in the spirit of \cite{dudley1987universal}. We moreover link these covering
 numbers with the correlation structure of the design.
 Section \ref{conclusion.section} concludes and Section \ref{proofs.section}
 contains proofs.

 \section{Notation and definitions} \label{definitions.section}
 
 \subsection{The compatibility constant}
  Let $S\subset \{ 1 , \ldots, p\}$ be an index set with cardinality $s$.
  We define for all $\beta \in \R^p$, 
 $$\beta_{S,j} := \beta_j {\rm l} \{ j \in S\} , \ j=1 , \ldots , p, \ \beta_{S^c} := \beta - \beta_S . $$
 
% Write $\Psi_S := \{ \psi_j \}_{j \in S} \cup \{ - \psi_j \}_{j \in S}$, and
% $\Psi_{S^c}:=  \{ \psi_j \}_{j \notin S} \cup \{ - \psi_j \}_{j \notin S} \cup \{0 \}$.
 Below, we present for constants $L>0$ the compatibility constant $\phi(L,S)$
 introduced in \cite{vandeG07}.
 For normalized $\psi_j$ (i.e., $\| \psi_j \|_n = 1 $ for all $j$), one
 can view $1- \phi^2 (1,S)/2$ as an $\ell_1$-version of the canonical correlation
 between the linear space spanned by the variables in $S$ on the one hand,
 and the linear space of the variables in $S^c$ on the other hand.
 Instead of all linear combinations with normalized $\ell_2$-norm,
 we now consider all linear combinations with normalized $\ell_1$-norm of the coefficients.
 For a geometric interpretation, we refer to Section \ref{measurecorr.section}.

 {\bf Definition} {\it The {\rm compatibility constant} is
 $$\phi^2 (L, S) := \min \{s  \| f_{\beta_S} - f_{\beta_{S^c}} \|_n^2   : \ \|\beta_S \|_1 =1 , \ 
 \| \beta_{S^c} \|_1 \le L  \} . $$ }

 The compatibility constant is closely related to (and never smaller than) the restricted
 eigenvalue as defined in \cite{bickel2009sal}, which is
 $$\phi_{\rm RE}^2 (L,S)= 
 \min \biggl \{ {  \| f_{\beta_S} - f_{\beta_{S^c}} \|_n^2 \over
 \| \beta_S \|_2^2 }   : \|\beta_{S^c} \|_1 \le L
 \| \beta_{S} \|_1 \biggr  \} . $$ 
 
See also \cite{koltch09a}, and see \cite{vdG:2009} for a discussion of the
relation between restricted eigenvalues and compatibility. 

\subsection{Projections}
As the ``true" $\beta^0$ is perhaps only approximately sparse,
 we will consider a sparse approximation. 
  The projection of $f^0 := f_{\beta^0}$ on the space spanned by the
 variables in $S$ is
 $${\rm f}_S := \arg \min_{f= f_{\beta_S}} \| f- f^0 \|_n . $$
 The coefficients of ${\rm f}_S$ are denoted by $b^S$, i.e., 
 $${\rm f}_S = f_{b^S} . $$ Note that ${\rm f}_S$ only has non-zero
 coefficients inside $S$, that is, $(b^S)_S = b^S$.

 \section{Main result}\label{main.section}
We let ${\cal T}_{\alpha}$ be the set
 $${\cal T}_{\alpha}:=  \biggl \{ \sup_{\beta} {  4| \epsilon^T f_{\beta} | /n\over
 \| f_{\beta} \|_n^{1- \alpha} \| \beta \|_1^{\alpha}  }  \le \lambda_0 \biggr \} . $$
 Here, $0 \le \alpha \le 1 $ and $\lambda_0>0$ are fixed constants.
 
 Note that on ${\cal T}_{\alpha}$,
 $$\sup_{\| \beta  \|_1 = 1 ,\ \| f_{\beta} 
 \|_n \le R } |\epsilon^T f_{\beta}   | /n  \le \lambda_0 R^{1-\alpha}/4  , $$
 i.e., we have a refinement of the dual norm bound described in Section
 \ref{intro.section}. 
 
 Note that for fixed $\lambda_0$ and for
 $\alpha < \tilde \alpha$, it holds that ${\cal T}_{\alpha} \subset {\cal T}_{\tilde \alpha}$.
 This is because by the triangle inequality
 $$\| f_{\beta } \|_n = \| \sum_{j}  \psi_j \beta_j \|_n \le \sum_{j}\| \psi_j \|_n | \beta_j |
 \le \| \beta \|_1 . $$
 We want to choose $\alpha$ preferably small, yet keep the probability of the
 set ${\cal T}_{\alpha}$ large. For $\alpha=1$, one has
 $${\cal T}_1 = \biggl \{ \max_{1 \le j \le p } 4 |\epsilon^T \psi_j |/n \le \lambda_0 
 \biggr \} , $$
 by the dual norm bound. Thus, e.g. when $\epsilon \sim {\cal N} (0, I)$,
 the probability $\PP ({\cal T}_1 ) $ of ${\cal T}_1 $ is large when $\lambda_0 \asymp \sqrt {\log p / n } $.  We detail in Section \ref{dudley.section} how
 one can lowerbound $\PP({\cal T}_{\alpha})$ for a proper value of $\alpha$
 depending on the design $\{ \psi_j \}$. Generally, the value for
 $\lambda_0$ will be of order $\sqrt {\log p / n }$, as in the case $\alpha=1$,
 or $\lambda \asymp \sqrt {\log n / n}$ or even
$ \lambda_0 \asymp 1/\sqrt {n}$. 
 
 The choice of the tuning parameter $\lambda$ depends on $\lambda_0$.
 The following technical lemma will be used:
 
\begin{lemma} \label{conjugate.lemma}
Let $0 \le \alpha \le 1 $ and let $a$, $b$ and $\lambda_0$ be positive
numbers. Then
$$ \lambda_0 a^{1- \alpha} b^{\alpha} \le {1 \over 2}  a^2 + \lambda b +
{1 \over 2}  \biggl ( { \lambda_0 \over \lambda^{\alpha} } \biggr )^{2 \over 1- \alpha} . $$
 Here, when $\alpha =1$, 
 $$
  \biggl ( { \lambda_0 \over \lambda^{\alpha} } \biggr )^{2 \over 1- \alpha}=
  \biggl ( { \lambda_0 \over \lambda } \biggr )^{\infty} :=
  \begin{cases} \infty & \lambda < \lambda_0  \cr
  1 & \lambda= \lambda_0  \cr 0 &  \lambda > \lambda_0  \cr  \end{cases}. $$
\end{lemma}

In the proof of the main result, Theorem \ref{main.theorem}, we invoke
Lemma \ref{conjugate.lemma} to handle the
``noise part"
$\epsilon^T f_{\beta}$ with
$\beta = \hat \beta - \beta^0$ (or actually
with $\beta^0$ replaced here by a sparse approximation).
On ${\cal T}_{\alpha}$,  it holds that 
$$4 | \epsilon^T f_{\beta} | /n\le{1 \over 2}  \| f_{\beta} \|_n^2   +  \lambda \| \beta \|_1 +
{1 \over 2}  \biggl ( { \lambda_0 \over \lambda^{\alpha} } \biggr )^{2 \over 1- \alpha} , $$
uniformly in $\beta \in \R^p$. In the right hand side
of this inequality, the  first term $ \| f_{\beta} \|_n^2 / 2 $ can be incorporated in the
risk and the second term
$ \lambda \| \beta \|_1$ will be overruled
by the penalty. Finally,  the third term $  ( { \lambda_0 / \lambda^{\alpha} }  )^{2 \over 1- \alpha} /2$ governs the choice of the tuning parameter $\lambda$. 

 We now come to the main result. We formulate it for
 an arbitrary index set $S$ partitioned in sets $S_1$ and $S_2$ in an arbitrary way.
We will elaborate on the choice of $S$
  in
 Remarks \ref{minimizeoverS} and  \ref{tradeoff.remark}.
  Corollaries \ref{classic.corollary} and \ref{slowrate.corollary} take for a given $S$ some
  special choices for the tuning parameter $\lambda$ and for the partition of $S$ into 
  $S_1 $ and $S_2$.
  
  Recall that ${\rm f}_S$ is the projection of $f^0= f_{\beta^0}$ and
  $b^S$ are the coefficients of ${\rm f}_S$. 
 
 \begin{theorem}\label{main.theorem}
 Let $S$ be an arbitrary index set, partitioned into two sets
 $S_1$ and $S_2$, i.e.\ $S= S_1 \cup S_2 $, $S_1 \cap S_2 = \emptyset$.
 Let $s_1$ be the cardinality of $S_1$.
Let ${\cal T}_{\alpha}$ be the set
 $${\cal T}_{\alpha} := \biggl \{ \sup_{\beta} 
 {4 | \epsilon^T f_{\beta}|/n \over \| f_{\beta}  \|_n^{1- \alpha}
 \| \beta \|_1^{\alpha} } \le \lambda_0 \biggr \} . $$
 Then on ${\cal T}_{\alpha}$, 
 $$\| \hat f - f^0 \|_n^2 + \lambda \| \hat \beta - b^S \|_1 \le
 { 56 \lambda^2 s_1 \over \phi^2 (6, S_1) } + {28 \over 3} \lambda \| ( b^S)_{S_2} \|_1 +
{7 \over 6}  \biggl ( { \lambda_0 \over \lambda^{\alpha} } \biggr )^{ 2 \over 1- \alpha } +
7  \| {\rm f}_S - f^0 \|_n^2 . $$
  \end{theorem}

  \begin{remark} \label{constants} 
  We did not attempt to optimize the constants we provided in Theorem \ref{main.theorem}. 
  \end{remark}
  
   \begin{remark} \label{minimizeoverS}
   Given a value of the tuning parameter $\lambda$, we can now define the estimation error
  using the variables in $S$ as
  $$ {\cal E} (S):= \min_{S_1 \subset S, \ S_2 = S \backslash S_1} 
   { 8 \lambda^2 s_1 \over \phi^2 (6, S_1) } + {4 \over 3} \lambda \| ( b^S)_{S_2} \|_1 .
 $$
 The oracle set $S_*$ is then the set which trades off estimation error and
 approximation error, i.e, the set $S_*$ that minimizes
 $${\cal E} (S) + \| {\rm f}_S - f^0 \|_n^2 . $$
 Note that $S_*$  depends on $\lambda$, say $S_*= S_* (\lambda)$.
 The best value for the tuning parameter $\lambda^*$ is then obtained by minimizing
 $$ {\cal E} ( S_* (\lambda)) + 
{1 \over 6}  \biggl ( { \lambda_0 \over \lambda^{\alpha} } \biggr )^{2 \over 1- \alpha} +\| {\rm f}_{S_*(\lambda)} - f^0 \|_n^2 . $$
  \end{remark}
  
    \begin{remark} \label{cross-validation} In practice, the tuning parameter
    $\lambda$ can be chosen by cross-validation. As this method tries to mimic
    minimization of the prediction error, it can be conjectured that one then arrives
    at rates at least a good as the ones we discuss here choosing  values of $\lambda$
    depending on the design, the (unknown) error distribution, and  the unknown sparsity.
    This is however not rigorously proven. 
     \end{remark}
  
  \begin{remark}\label{generalconjugates}
  We have restricted ourselves to improvements of the
  dual norm bound of the form given by sets ${\cal T}_{\alpha}$. The situation can be
  generalized by considering sets of the form
  $$ \biggl \{ \sup_{\beta} {4 |\epsilon^T f_{\beta} |/n \over
  G^{-1} ( \| f_{\beta } \|_n / \| \beta \|_1 ) \| \beta \|_1 }  \le \lambda_0 \biggr \} , $$
  where $G$ is a given increasing convex function with $G(0)=0$. 
  \end{remark}

  \begin{corollary} \label{classic.corollary} $ $ \\
  a) If we take $S_2= \emptyset$, we have $S_1= S$,
  and $s_1 = |S|=:s $. This is a good choice when the compatibility constants
  are large for all subsets of $S$. 
  With the choice
  $$\lambda^2 \asymp \lambda_0^2 \biggl ( { \phi^2 (6, S) \over s } \biggr )^{1- \alpha} , $$
  we get on ${\cal T}_{\alpha} $,
  $$\| \hat f - f^0 \|_n^2 + \lambda \| \hat \beta - b^S \|_1 = {\mathcal O} \left (  \lambda_0^2
\biggl  (  {s \over \phi^2 ( 6,S) } \biggr )^{\alpha} + \| {\rm f}_S - f^0 \|_n^2 \right ) . $$
Recall that  the dual norm bound has $\alpha=1$. With $\lambda_0 \asymp
\sqrt {\log p / n } $ we then arrive at the ``usual" oracle inequality as
provided by, among others,  \cite{bickel2009sal}, \cite{Bunea:06},  \cite{Bunea:07a}, \cite{Bunea:07b},  \cite {koltch09a}
 \cite {vandeG07}, \cite {vandeG08}. 
When $\alpha< 1$, the compatibility constant may be very small,
as the design is highly correlated. The effect is however somewhat tempered by
the power $\alpha$ in the bound.\\
b) More generally, let 
$$\lambda^2 \asymp \lambda_0^2 \biggl ( { \phi^2 (6, S_1) \over s_1 } \biggr )^{1- \alpha} , $$
Then on ${\cal T}_{\alpha}$, 
$$\| \hat f - f^0 \|_n^2 + \lambda \| \hat \beta - b^S \|_1 $$ $$= {\mathcal O} \left (  \lambda_0^2
\biggl  (  {s_1 \over \phi^2 ( 6,S_1) } \biggr )^{\alpha} +
\lambda_0 \biggl ( {\phi^2 (6, S_1) \over s_1 } \biggr )^{1-\alpha \over 2} 
\| (b^S)_{S_2} \|_1 +   \| {\rm f}_S - f^0 \|_n^2 \right ) . $$
\end{corollary}

\begin{corollary}\label{slowrate.corollary} $ $ \\
a) With the choice $S_1 = \emptyset$,
the result does not involve the compatibility constant. 
This may be desirable when the design is highly correlated.
The result then corresponds to what is sometimes called ``slow rates", although we will
see that when $\alpha < 1$, the rates can still be much faster than $1/\sqrt{n}$.
When $\alpha =1$, we must take
$\lambda > \lambda_0$ (due to the term $(\lambda_0 / \lambda^{\alpha})^{2 \over 1- \alpha} $). When 
$\alpha< 1$, we choose
$$\lambda \asymp \lambda_0^{2 \over 1+ \alpha} \| b^S \|_1^{-{1-\alpha \over 1+ \alpha} } .$$
  We get on ${\cal T}_{\alpha}$,
  $$\| \hat f - f^0 \|_n^2+\lambda \| \hat \beta - b^S \|_1  = {\mathcal O} \left ( 
  \lambda_0^{2 \over 1+ \alpha} \| b^S \|_1^{2 \alpha  \over 1+ \alpha} + \| {\rm f}_S - f^0 \|_n^2 \right ) . $$\\
  b) More generally, let
  $$\lambda \asymp \lambda_0^{2 \over 1+ \alpha} \| (b^S)_{S_2} \|_1^{-{1-\alpha \over 1+ \alpha} } .$$
  Then on ${\cal T}_{\alpha} $,
  $$\| \hat f - f^0 \|_n^2+\lambda \| \hat \beta - b^S \|_1  $$ $$= {\mathcal O} \left ( 
  {\lambda_0^{4 \over 1+\alpha} \over \| (b^S)_{S_2} \|_1^{2(1-\alpha) \over
  1+ \alpha} } { s_1 \over \phi^2 (6, S_1) } + 
  \lambda_0^{2 \over 1+ \alpha} \| (b^S)_{S_2}  \|_1^{2 \alpha  \over 1+ \alpha} + \| {\rm f}_S - f^0 \|_n^2 \right ) . $$
\end{corollary}

\begin{remark}\label{tradeoff.remark}

Note that by taking $S_1$ smaller, the value of $s_1/ \phi^2 (6, S_1)$
will not
increase, but on the other hand, the value of $\| (b^S)_{S_2 } \|_1$ will become
larger. Thus, the best rate will emerge if we trade off these two effects.
Indeed, suppose that for some $S_1$ 
$$ \lambda_0^{2 \over 1+ \alpha}  { s_1 \over  \phi^2 (S_1) } \asymp 
\| (b^S)_{S_2} \|_1^{2 \over 1+ \alpha} . $$
Then on ${\cal T}_{\alpha}$, for 
$$\lambda \asymp \lambda_0 \biggl ( {s_1^2  \over  \phi^2 (6, S_1)} \biggr )^{(1-
\alpha)/2} \asymp
\lambda_0^{2 \over 1+ \alpha} \|( b^S)_{S_2} \|_1^{-{1-\alpha \over 1+ \alpha} } , $$ we have
$$\| \hat f - f^0 \|_n^2+\lambda \| \hat \beta - b^S \|_1  = 
{\mathcal O} \left (  \lambda_0^2
\biggl  (  {s_1 \over \phi^2 ( 6,S_1) } \biggr )^{\alpha} + \| {\rm f}_S - f^0 \|_n^2 \right )
$$ $$= {\mathcal O} \left ( 
  \lambda_0^{2 \over 1+ \alpha} \| (b^S )_{S_2}\|_1^{2 \alpha  \over 1+ \alpha} + \| {\rm f}_S - f^0 \|_n^2 \right ) . $$
  In particular for the case $\alpha < 1$, 
  it is however not clear when such a trade-off is possible. It may well be that
  for any $S_1 $, 
$s_1 / \phi^2 (6, S_1)$ either heavily dominates or is heavily dominated by the $\ell_1$-part
$\| (b^S)_{S_2} \|_1$. See Section \ref{measurecorr.section} for a further discussion. 
  
  %  Assuming that for all $S$, we have such a set $S_1 \subset S$ which trades off the two effects, 
%  minimizing the above bounds over all $S$ 
%  leads to a bound of the order
%  ${\cal E} (S_* (\lambda^*) ) +
%  ( { \lambda_0 / (\lambda^*)^{\alpha} }  )^{2 \over 1- \alpha}+
%  \| {\rm f}_{S_* (\lambda^*)} - f^0 \|_n^2 $,
%  as in Remark \ref{minimizeoverS}
 \end{remark}

 \section{Improving the dual norm bound} \label{dudley.section}
 
 In this section, we provide probability
 bounds for the set ${\cal T}_{\alpha}$ introduced in Section
 \ref{main.section}.
 The results follow from empirical process theory, see e.g.  and \cite{vandeG} and 
 \cite{vanderVaart:96}. Theorem \ref{polynomialentropy.theorem}
 is taken from \cite{BvdG2011}. 
 
 {\bf Definition} {\it Let ${\cal F}$ be a class of real-valued functions on ${\cal X}$.
 Endow ${\cal F}$ with norm $\| \cdot \|_n$. Let $\delta >0$ be some radius.
 A {\rm $\delta$-packing set} is a set of functions in ${\cal F}$ that
 are each at least $\delta$ apart.
  A {\rm $\delta$-covering set}  is a set of functions
 $\{ \phi_1 , \ldots , \phi_N \}$, such that
 $$\sup_{f \in \cal F} \min_{k=1 , \ldots , N} \| f - \phi_k \|_n \le \delta . $$
 The {\rm $\delta$-covering number} $N(\delta , {\cal F} , \| \cdot \|_n )$ of ${\cal F}$ is the minimum size of a $\delta$-covering set. The {\rm entropy} of ${\cal F}$ is
$H( \cdot , {\cal F} , \| \cdot \|_n )=  \log N( \cdot , {\cal F} , \| \cdot \|_n ) $. }

It is easy to see that $N(\delta , {\cal F} , \| \cdot \|_n )$ can be
bounded by the size of a maximal $\delta$-packing set. 

We assume the errors are sub-Gaussian, that is,  for some positive constants $K$ and
$\sigma_0$, 
 \begin{equation}\label{subgaussian.assumption}
 K^2 (\EE \exp[ \epsilon_i^2 / K^2 ] -1)\le \sigma_0^2 , \ i=1 , \ldots , n . 
 \end{equation}
 
 The following theorem is Corollary 14.6 in \cite{BvdG2011}.
 It is in the spirit of a weighted concentration inequality, and uses the notation
 $$x_+ := \max\{x, 0 \} . $$

 \begin{theorem} \label{polynomialentropy.theorem}
 Assume (\ref{subgaussian.assumption}).
 Let ${\cal F}$ be a class of functions with $\| f \|_n \le 1$ for all $f \in {\cal F}$, and with,
 for some $0 < \alpha < 1$ and some constant $A $, 
 $$\log\biggl  (1+ 2N(\delta , {\cal F} , \| \cdot \|_n ) \biggr ) \le \left ( { A\over \delta} \right )^{2\alpha } , \ 0 < \delta \le 1 . $$
% Then
% $$a_0(R) \le {1 \over 2}  A_{\nu}^{\nu} R^{1-\nu } \sum_{s=1}^{\infty}  2^{-s(1-\nu ) } =
% {1 \over 2}  A_{\nu}^{\nu} R^{1-\nu} (2^{1-\nu} -1 )^{-1} := a (R). $$
% Moreover
% $$\sum_{s=0}^{\infty} \exp\biggl [ -{ 2 a^2 (2^{-s}) \over 2^{-2s}  }  \biggr ]
% $$ $$ = \sum_{s=0}^{\infty} \exp\biggl  [ - { A_{\nu} ^{2 \nu} 2^{2\nu s} \over 
% 2 ( 2^{1-\nu} -1 )^2 }   \biggr ] 
%  \le \sum_{s=0}^{\infty} \exp\biggl  [ - { A_{\nu}^{2 \nu} \nu (s+1) \over 
% 2 ( 2^{1-\nu} -1 )^2 }   \biggr ]  $$ 
Define
$$B:=
 \exp \biggl [ { A^{ 2 \alpha } \alpha \over 2 ( 2^{1-\alpha} -1 )^2} \biggr ] - 1 ,$$
and 
$$ K_0 := 3 \times 2^5
\sqrt {
K^2+\sigma_0^2}
 . $$

 It holds that
 $$\EE \exp \left [ \sup_{f \in {\cal F} } \biggl [ \left (
 { |\epsilon^Tf   |/ \sqrt {n} \over \| f \|_n K_0 } -  { A^{\alpha} \| f \|_n^{-\alpha} \over  2^{1-\alpha} -1 }
 \right )_+ 
  \biggr ]^2 \right ]  \le 1 +   2 /B. $$

     \end{theorem} 
 
 \begin{corollary}\label{probineq.corollary}
 Assume the conditions of Theorem \ref{polynomialentropy.theorem}.
  Chebyshev's inequality shows that for all $t>0$,
   $$\PP \biggl(\exists \ f \in {\cal F} : \   | \epsilon^T f |/ \sqrt n  \ge K_0 A^{\alpha} 
   \| f\|_n^{1-\alpha}
   (2^{1-\alpha} -1 )^{-1} + K_0 \| f \|_n t  \biggr ) $$
   $$ \le   \exp [ - t^2  ]  (1+ 2 /B) . $$
  \end{corollary}
 
  \begin{corollary} \label{final.corollary} Consider now linear functions
  $$f_{\beta} := \sum_{j=1}^p \psi_j  \beta_j, \ \beta \in \R^p , $$
  where $\| \psi_j \|_n \le 1 $. Then 
  $$\| f_{\beta} \|_n \le \| \beta \|_1 . $$
  Hence, $\{ f_{\beta} / \| \beta \|_1 : \beta \in \R^p \}$ is a class of functions
  with $\|\cdot \|_n$-norm bounded by $1$. Suppose now
  $$\log \biggl (1+2N( \delta , \{ f_{\beta}: \ \| \beta \|_1 =1 \} , \| \cdot \|_n ) \biggr )\le 
 \left  ( { A\over \delta} \right )^{2\alpha } , \ 0 < \delta \le 1 . $$
Under the sub-Gaussianity condition (\ref{subgaussian.assumption}), we then have
for all $t >0$ and for 
$$\lambda_0 =  {4 K_0 \over \sqrt n}  \biggl ( { A^{\alpha}   \over 2^{1-\alpha} -1 }  +t  \biggr ) ,$$
the lower bound
$$\PP ({\cal T}_{\alpha}  ) 
    \ge1-   \exp [ - t^2 ]  (1+ 2 /B) . $$

 \end{corollary}

 \section{Compatibility, eigenvalues, entropy and correlations}\label{measurecorr.section}
 
 We study the set
 $${\cal F} := \{ f_{\beta}: \| \beta \|_1 = 1 \} . $$
  It is considered as subset of
 $L_2(Q_n)$, where $Q_n := \sum_{i=1}^n \delta_{x_i}/n $.
 The $L_2 (Q_n)$-norm  is $\| \cdot \|_n $. 
 
 \subsection{Geometric interpretation of the compatibility constant}
 \label{geometric.section}
\begin{figure} % figuur 1
\centerline{
  \includegraphics[height=1.2in, width=2in]{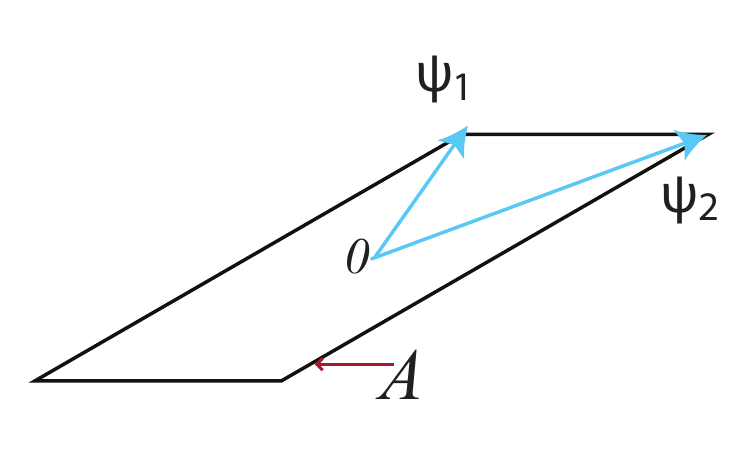}
  \includegraphics[height=1.2in, width=2in]{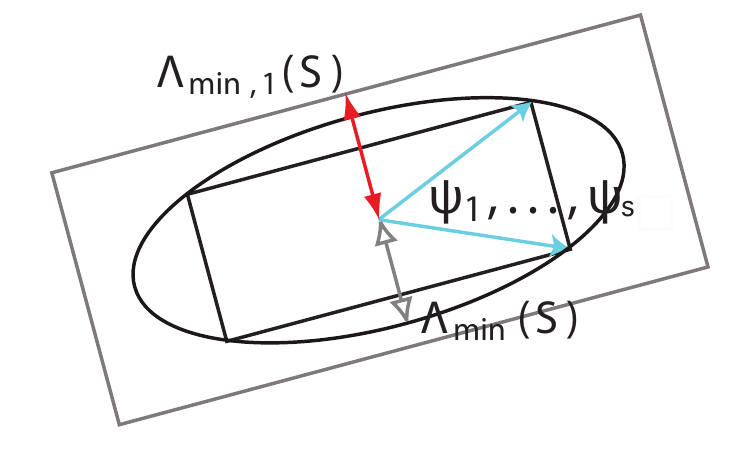}
}
\caption[]{Left panel: the set $A= \{ f_{\beta_S}: \| \beta_S \|_1 = 1 \} $. Right panel: $\ell_1$- and $\ell_2$-eigenvalues.}
\label{Figure1}
\end{figure}

  We first look at the minimal
 $\ell_1$-eigenvalue
 $$\Lambda_{ {\rm min} , 1 }^2 (S ):=
 \min\biggl  \{ s \beta_S^T \hat \Sigma \beta_S: \ \| \beta_S \|_1 = 1 \biggr \} $$
 as introduced in \cite{BvdG2011}.
 Note that $\Lambda_{ {\rm min} , 1 } (S )/ \sqrt {s}$ is the minimal distance between
 any point $f_{\beta_{S}}$ with $\| \beta_S \|_1 = 1 $ and the point $\{ 0 \}$.
 We tacitly assume that the $\{ \psi_j \}_{j \in S}$ are linearly independent.
 The set $\{ f_{\beta_S} : \ \| \beta_S \|_1 =1 \}$ is then an $\ell_1$-version
 of a sphere: it is the boundary of the convex hull of $\{ \psi_j \}_{j \in S} \cup
 \{ - \psi_j \}_{j \in S} $ in $s$-dimensional space with $\{ 0 \}$ in
 its ``center". It is a parallelogram when $s=2$ (see Figure \ref{Figure1}) and then a rectangle when the $\psi_j$, $j \in S$, have equal length. 

\begin{figure} % figuur 1
\centerline{
   \includegraphics[height=1.35in, width=2.25in]{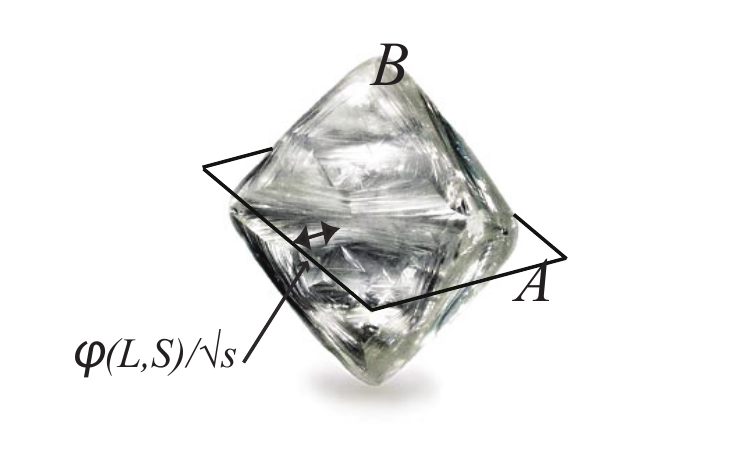}
}
\caption[]{The compatibility constant}
\label{Figure2}
\end{figure}

 Let $\hat \Sigma_S$ be the Gram matrix of the variables in $S$
and $\Lambda_{\rm min}^2 ( S)$ be the minimal ($\ell_2$-)eigenvalue
 of the matrix $\hat \Sigma_S $:
 $$\Lambda_{ {\rm min}  }^2 (S ):=
 \min\biggl  \{ \beta_S^T \hat \Sigma \beta_S: \ \| \beta_S \|_2 = 1 \biggr \} . $$
Then 
 $$\Lambda_{ {\rm min} , 1 }^2  (S)
 \ge \Lambda_{\rm min}^2 (S) \ge \Lambda_{ {\rm min} , 1 }^2  (S) / s , $$
  One can construct examples where $\Lambda_{\rm min}^2 (S)$
 is as small as $3 / (s-2)$ ( $ s > 2$) and 
  $\Lambda_{{\rm min},1}^2 (S)$ is at least $1/2$
  (see \cite{vdG:2009}), that is,  they can differ by the maximal
  amount $s$ in order of magnitude. See also Figure \ref{Figure1} which is to be understood as representing a case $s>2$.  Thus,
  minimal $\ell_1$-eigenvalues can be much larger than minimal
  ($\ell_2$-)eigenvalues. 
The normalized compatibility constant $\phi(L,S)/\sqrt {s}$ is the minimal distance  between the sets
 $A:=\{f_{\beta_S} : \ \| \beta_S \|_1 =1 \} $ and
 $B:= \{  f_{\beta_{S^c}} : \ \| \beta_{S^c} \|_1 \le L \} $, that is,
 $${ \phi(L,S) \over \sqrt s} = \min \biggl \{ \| a-b \|_n: \ a \in A , \ b \in B \biggr \} . $$
 See Figure \ref{Figure2} for an impression of the situation.
 Observe that $A$ is the  boundary of the convex hull of $\{+ \psi_j \}_{j \in S} \cup \{- \psi_j \}_{j \in S} $, and $B$ is 
 the convex hull of $\{+ \psi_j \}_{j \in S^c} \cup \{- \psi_j \}_{j \in S^c} $ including
 its interior, 
 blown up with a factor $L$ (typically, the 
 $\{ \psi_j \}_{j \in S^c}$ form a linearly dependent system in $\R^n$).
 Furthermore, since $\{ 0 \}  \in B$
 $$\phi(L,S) \le \Lambda_{ {\rm min} , 1 } (S) . $$
 This shows that when $\ell_1$-eigenvalues are small,
 the compatibility constant is necessarily also small. 
 Small $\ell_2$-eigenvalues may have less of this effect.

  \subsection{Eigenvalues and entropy} \label{eigenvalues-entropy.section}
  We now let
  $$\hat \Sigma = E \Omega^2 E^T  $$
  be the spectral decomposition of the Gram matrix $\hat \Sigma$,
  $E$ being the matrix of eigenvectors, 
  ($E^T E = E E^T = I$) and $\Omega^2= {\rm diag} ( \omega_1^2 , \cdots ,
  \omega_p^2 ) $ the matrix of ($\ell_2$-)eigenvalues. We assume
  they are in decreasing order: $\omega_1^2 \ge \cdots \ge \omega_p^2 $. 
  
  \begin{lemma}\label{eigenvalue-entropy} Suppose that for some
  strictly decreasing function $V$
  $$\omega_{j+1}^2 \le V^2(j), \ j=1 , \ldots , p . $$
  Then for all $\delta >0$, 
  $$H ( 2 \delta , \{ f_{\beta}: \ \| \beta \|_1 = 1 \} , \| \cdot \|_n ) \le
 V^{-1} ( \delta)  \log \biggl ( { 3 \over \delta } \biggr ) . $$
   \end{lemma}

  \begin{example}\label{polydecrease}
  Suppose that for some positive constants $m $ and $C$
  $$\omega_j \le { C \over j^m}, \ j=1 , \ldots , p .  $$
  Then by Lemma \ref{eigenvalue-entropy}, 
  $$H (2 \delta, \{ f_{\beta} : \ \| \beta \|_1 = 1 \} , \| \cdot \|_n ) \le
  \biggl ( {C \over \delta } \biggr )^{1 \over m} \log  \biggl ({3 \over \delta }  \biggr ). $$
  For $\delta \ge1/n$ (say) we therefore have
   $$H ( \delta, \{ f_{\beta} : \ \| \beta \|_1 = 1 \} , \| \cdot \|_n ) \le
  \biggl ( {2 C \over \delta } \biggr )^{1 \over m} \log (6n). $$

  When $m > 1/2$, one can use a minor
  generalization of Corollary \ref{final.corollary}, where the entropy bound
  is only required for values of $\delta > 1/n$. One then takes
  $$\alpha = {1 \over 2m} , \ A= (C_m^2 \log (n) )^{1 \over 2\alpha},  $$
where $C_m$ is a constant depending on $m$ and $C$.
  Then the value of $\lambda_0$ defined there becomes
  $$\lambda_0 = {4 K_0 \over \sqrt n} \biggl ( 
  { C_m \sqrt {\log (n) } \over
  2^{1- {1 \over 2m}} -1 } + t \biggr ) $$
  which is for fixed $m$ and $K_0$, and a fixed (large) $t$, of order
  $\sqrt {\log n / n } $. 
  
  \end{example}
  
  \subsection{Entropy based on coverings of $\{ \psi_j \} $}
  \label{entropy-covering.section}
  
  We can consider
  $ \{ f_{\beta}: \| \beta \|_1 = 1 \} $
 as a subset of
 $${\rm conv} ( \{ \pm \psi_j \}) , $$
 where $\{ \pm \psi_j \} := \{ \psi_j \} \cup \{ - \psi_j \} $, and
 ${\rm conv} (\{ \pm \psi_j \} ) $ is its convex hull. 
 Infact, if the $\{ \psi_j \}$ form a linearly dependent system in $\R^n$, ${\cal F}$
 is exactly equal to ${\rm conv} (\{  \pm \psi_j\})$. 

  The paper \cite{dudley1987universal} gives a bound for the entropy of a convex hull
   for the case where the $u$-covering number of the extreme points 
 is a polynomial in $1/u$. This result can also be found in
   \cite{pollard1990empirical}.
     There is a redundant $\log$-term in these entropy bounds,
   see \cite{Ball:90} and  \cite{vanderVaart:96}, but removing this $\log$-term may
   result in very large constants, depending on the
   dimension $W$ as given in Example \ref{polycovering.example}
   (see \cite{BvdG2011} for some explicit constants). 
   This means that when the dimension
   $W$ of the extreme points is large (growing with $n$ say),
   the simple bound with $\log$-term we provide below  in Lemma
   \ref{withlog} may be better than the more
   involved ones.

     We give a
     bound for the entropy of ${\cal F}$ by balancing the $u$-covering number 
     of $\{\psi_j \}$ and the
squared radius $u^2$. The result is
as in  \cite{pollard1990empirical}, with only new element its extension
   to general covering numbers (i.e., not only polynomial ones).
Lemma \ref{withlog} and its proof can be found in \cite{BvdG2011}.

\begin{lemma} \label{withlog} 
Let 
$$N(u) := N( u, \{  \psi_j \}, \| \cdot \|_n ) , \ u > 0 . $$
We have 
$$H\biggl (\delta , \{ f_{\beta}: \ \| \beta \|_1 =1 \}  , \| \cdot \|_n\biggr  ) 
$$ $$\le
            \min_{0<u <1}   6
  \left ( N(u) +{  6 u^2 \over \delta^2 }  \right )  \log \biggl ( 2 \biggl ( { 8 + \delta \over
  \delta} \biggr  ) N(\delta ) \biggr ) 
 .$$
   \end{lemma}

 \begin{example}\label{polycovering.example}
 In this example, we assume the $u$-covering numbers of $\{ \psi_j \}$ are
 bounded by a polynomial in $u$. That is, we 
 suppose that for some positive constants $W$ and $C$, 
 $$N(u, \{\psi_j \} , \| \cdot \|_n ) \le \biggl ({ C \over u}\biggr )^W ,  u > 0 . $$
 The constant $W$ can be thought of as the dimension of $\{ \psi_j \}$.
 By Lemma \ref{withlog}, we can choose
 $$\alpha = {W \over 2+W}, \ A= ( C_W^2 \log (n) )^{1 \over 2 \alpha } , $$
 and we get, as  in Example \ref{polydecrease},
 $$\lambda_0 = { 4 K_0 \over \sqrt n} \biggl ( 
  { C_W \sqrt {\log (n) } \over
  2^{{2 \over 2 +W}} -1 } + t \biggr ) .$$
 
 \end{example}

  A refined analysis of the relation between compatibility
  constants, covering numbers and entropy is still to be carried out.
  We confine ourselves here to the following, rather trivial, observation
  (without proof).
  
  \begin{lemma}\label{compa.lemma} 
  Consider normalized design:
  $\| \psi_j \|_n = 1 $ $\forall \ j$.
  Let
$\{ \psi_{j_1} , \ldots , \psi_{j_N} \} 
$ be a maximal $u$-packing set of $\{ \psi_j \} $.
%\{ f_{\beta} : \| \beta \|_1 \le 1 \}$.
%$M= M(\rho)$. Then for $S = $ $\supset ? \{ j_1 , \ldots , j_M \} $, 
%$$ \Lambda_{\rm min}^2 (S) > 1-\rho 
%$$ and for any
%$L \ge 1 $,
%$$\phi^2 (L,S) \le 1-\rho . $$
%??
Then for any $S\supset \{ j_1 , \ldots , j_N \}$, $S \not= \{ 1, \ldots , p  \}$,
and any 
$L \ge 1 $,
$$\phi^2 (L,S) \le s u^2 . $$
\end{lemma}

One may argue that as $u$-packing sets are approximations
of the original design $\{ \psi_j \}$ with fewer covariables, 
they are good candidates for the sparsity set $S_1$
used in Theorem \ref{main.theorem}. Lemma \ref{compa.lemma} however
shows that such sparsity sets will have very small compatibility constants.

\subsection{Decorrelation numbers}
 Decorrelation numbers are closely related to packing numbers.
 First, define the inner product
 $$\rho( \phi , \tilde \phi ) := \phi^T \tilde \phi / n . $$
 Note that $\Sigma_{j,k} = \rho(\psi_j , \psi_k ) $ and that in the case of standardized design
 (i.e. $ \sum_{i=1}^n \psi_j (x_i) =0 $
 and $\| \psi_j \|_n = 1 $ $\forall \ j$),
 the inner product $\rho(\psi_j , \psi_k ) $ is for $j \not= k$ the (empirical)
 correlation between $\psi_j$ and $\psi_k$.

  {\bf Definition} {\it For $\rho>0$, the {\rm $\rho$-decorrelation number}
  $M(\rho)$ is the largest value of $M$ such that
  there exists 
 $\{ \phi_1 , \ldots , \phi_M \} \subset \{ \pm \psi_j \} $
 with $|\rho (\phi_j , \phi_k )| < \rho $ for all $j \not= k $.}

 Hence, if the $\rho$-decorrelation number is small, then there
 are many large correlations, i.e., then the design
 is highly correlated.
 
 It is clear that when $\| \psi_j \|_n = \| \psi_k \|_n = 1$, it holds that
 $$\| \psi_j - \psi_k \|_n^2 = 2(1- \rho ( \psi_j , \psi_k ) ) . $$
 In other words, small correlations correspond to covariables
 that are near to each other. This can be translated into covering number as shown in
 Lemma \ref{measurecorr.lemma}. Its proof is straightforward and omitted.

   \begin{lemma} \label{measurecorr.lemma}
Consider normalized design:
  $\| \psi_j \|_n = 1 $ $\forall \ j$.
For all $ 0<u  <1$, 
$$N( \sqrt 2  u, \{\pm \psi_j \} , \| \cdot \|_n ) \le M(1- u^2). $$
\end{lemma}

\section{Conclusion}\label{conclusion.section}
We have combined results for the prediction error
of the Lasso with both compatibility conditions and entropy conditions.
Small entropies of $\{ f_{\beta}: \ \| \beta \|_1 = 1\}$
correspond to highly correlated design and possibly to small
compatibility constants. Our analysis shows that
small entropies allow for a smaller choice of the tuning parameter and possibly
for a compensation of small compatibility constants. This means that
the Lasso enjoys good prediction error properties, even in the case
where the design is highly correlated. 

\section{Proofs}\label{proofs.section} 
$$ $$
{\bf Proof of Lemma \ref{conjugate.lemma}.} 
We use that for positive $u$ and $v$ and for ${\rm p} \ge 1$,
${\rm q} \ge 1$, $1/{\rm p} + 1 / {\rm q} =1 $, the conjugate inequality
$$uv \le u^{\rm p}  / {\rm p}+ v^{\rm q} /  {\rm q} $$
holds.
Taking ${\rm p} = 1/(1- \alpha)$ and replacing $u$ by $u^{1- \alpha} $
gives
$$u^{1- \alpha} v \le { 1 - \alpha \over 2} u^2 + { 1+ \alpha \over 2}
v^{2 \over 1+ \alpha }  . $$
With ${\rm p} = (1+ \alpha )/ (2 \alpha)$, and replacing
$u$ by $u^{2 \alpha \over 1+ \alpha} $, we get
$$u^{2 \alpha \over 1+ \alpha} v \le { 2 \alpha \over
1+ \alpha} u+ { 1- \alpha \over 1 + \alpha } v^{1 + \alpha \over 1- \alpha } . $$
Thus,
$$\lambda_0 a^{1-\alpha} b^{\alpha} \le { 1 - \alpha \over 2} a^2+
{ 1+ \alpha \over 2}
\biggl ( \lambda_0  b^{\alpha} \biggr )^{2 \over 1+ \alpha } $$
$$ \le { a^2 \over 2}  + { 1+ \alpha \over 2}(\lambda b)^{2 \alpha \over 1+ \alpha } 
 \biggl ({ \lambda_0 \over \lambda^{\alpha} } \biggl )^{2 \over 1+\alpha} 
 $$
$$ \le{  a^2 \over 2} + { 1+ \alpha \over 2} \left (
 { 2 \alpha \over
1+ \alpha} \lambda b + { 1- \alpha \over 1 + \alpha }
 \biggl ({ \lambda_0 \over \lambda^{\alpha} } \biggl )^{2 \over 1-\alpha}
 \right ) $$
 $$ \le { a^2 \over 2} + \lambda b + {1 \over 2}  \biggl ({ \lambda_0 \over \lambda^{\alpha} } \biggl )^{2 \over 1-\alpha} . $$
 \hfill $\sqcup \mkern -12mu \sqcap$
 
   {\bf Proof of Theorem \ref{main.theorem}.} 
 Since
 $$\| {\bf Y} - \hat f \|_2^2 / n + \lambda \| \hat \beta \|_1 \le 
 \| {\bf Y} - {\rm f}_S \|_2^2 / n + \lambda \| b^S \|_1 , $$
 we have
 the Basic Inequality
  $$\| \hat f - f^0 \|_n^2 +  \lambda \| \hat \beta \|_1 \le
  2 \epsilon^T ( \hat f - {\rm f}_S )/n +  \lambda \| b^S \|_1 + \| {\rm f}_S - f^0 \|_n^2 . $$
  Hence, on ${\cal T}_{\alpha}$,
$$\| \hat f - f^0 \|_n^2 +  \lambda \| \hat \beta \|_1 \le
 \lambda_0 \| \hat f - {\rm f}_S \|_n^{1- \alpha} 
\| \hat \beta - b^S \|_1^{\alpha} /2 +  \lambda \| b^S \|_1 + \| {\rm f}_S - f^0 \|_n^2 . $$
Apply Lemma \ref{conjugate.lemma} to find
$$\| \hat f - f^0 \|_n^2 +  \lambda \| \hat \beta \|_1 $$ $$\le{1 \over 4} 
\| \hat f - {\rm f}_S \|_n^2  +{1 \over 2}  \lambda \| \hat \beta - b^S \|_1 +{1 \over 4}
\biggl ( {\lambda_0 \over  \lambda^{\alpha} }\biggr )^{2 \over 1- \alpha } 
 +  \lambda \| b^S \|_1 + \| {\rm f}_S - f^0 \|_n^2 . $$
 $$\le {1 \over 2} 
\| \hat f - f^0 \|_n^2  +{1 \over 2}  \lambda \| \hat \beta - b^S \|_1 +{1 \over 4}
\biggl ( {\lambda_0 \over  \lambda^{\alpha} }\biggr )^{2 \over 1- \alpha } 
 +  \lambda \| b^S \|_1 + {3 \over 2 } \| {\rm f}_S - f^0 \|_n^2 . $$
 Thus, we get on ${\cal T}_{\alpha}$,
 $$\| \hat f - f^0 \|_n^2 +  2 \lambda \| \hat \beta \|_1\le 
 \lambda \| \hat \beta - b^S \|_1 +   2  \lambda \| b^S \|_1 + {1 \over 2}
\biggl ( {\lambda_0 \over  \lambda^{\alpha} }\biggr )^{2 \over 1- \alpha } 
 + 3   \| {\rm f}_S - f^0 \|_n^2 . $$
 Defining $S_3 := S^c$, we rewrite this to
 $$\| \hat f - f^0 \|_n^2 +  2 \lambda \| \hat \beta_{S_2 \cup S_3}  \|_1 $$ $$\le 
 \lambda \| \hat \beta_{S_1} - (b^S)_{S_1}  \|_1 
 +  \lambda \| \hat \beta_{S_2} - (b^S)_{S_2} \|_1 +
  \lambda \| \hat \beta_{S_3 } \|_1 +   2   \lambda \| (b^S)_{S_1} \|_1 
 - 2 \lambda \| \hat \beta_{S_1} \|_1 $$ $$ + 2 \lambda \| (b^S)_{S_2} \|_1 + {1 \over 2}
\biggl ( {\lambda_0 \over  \lambda^{\alpha} }\biggr )^{2 \over 1- \alpha } 
 + 3   \| {\rm f}_S - f^0 \|_n^2 $$
 $$ \le 3 \lambda \| \hat \beta_{S_1} - (b^S)_{S_1}  \|_1  +
  \lambda \| \hat \beta_{S_2 \cup S_3  } \|_1  + 3 \lambda \| (b^S)_{S_2} \|_1 + {1 \over 2}
\biggl ( {\lambda_0 \over  \lambda^{\alpha} }\biggr )^{2 \over 1- \alpha } 
 + 3   \| {\rm f}_S - f^0 \|_n^2 .$$
 Moving the term $\lambda \| \hat \beta_{S_2 \cup S_3} \|_1$
 to the left hand side, and applying a triangle inequality, we obtain
 $$\| \hat f - f^0 \|_n^2 +   \lambda \| \hat \beta_{S_2 \cup S_3} -(b^S )_{S_2}  \|_1 $$
$$ \le \underbrace{3 \lambda \| \hat \beta_{S_1} - (b^S)_{S_1}  \|_1  }_
{:= I} 
   + \underbrace{4  \lambda \| (b^S)_{S_2} \|_1 + {1 \over 2}
\biggl ( {\lambda_0 \over  \lambda^{\alpha} }\biggr )^{2 \over 1- \alpha } 
 + 3   \| {\rm f}_S - f^0 \|_n^2}_{:= II} .$$
 
 {\bf Case i.} If $I \ge II$, we arrive at
 $$\| \hat f - f^0 \|_n^2 +   \lambda \| \hat \beta_{S_2 \cup S_3} -(b^S)_{S_2} \|_1 \le
 6  \lambda \| \hat \beta_{S_1} - (b^S)_{S_1}  \|_1 . $$
We first add
 add a term $\lambda \| \hat \beta_{S_1} - (b^S)_{S_1} \|_1 $ to the left and
 right hand side and then apply the compatibility condition to $\hat \beta - b^S $, to get
  $$\| \hat f - f^0 \|_n^2 +   \lambda \| \hat \beta -b^S \|_1 \le
 { 7  \lambda  \sqrt {s_1}  \over
 \phi(6, S_1) } \| \hat f - {\rm f}_S \|_n   $$
 $$ \le{1 \over 2}  \| \hat f - f^0 \|_n^2 +{7 \over 2}  \| {\rm f}_S - f_0 \|_n^2 + 
  {56 \lambda^2 s_1 \over 2 \phi^2 (6, S_1)} . $$
  Here we used the decoupling device
  $$2xy \le bx^2 + y^2 / b \ \forall \ x,y \in R, \ b > 0 . $$
  So then
  $$\| \hat f - f^0 \|_n^2 +   2 \lambda \| \hat \beta -b^S \|_1 \le
  { 56 \lambda^2   s_1 \over
 \phi^2(6, S_1) }  + 7 \| {\rm f}_S - f^0\|_n^2 .   $$
 
 {\bf Case ii.} If $I < II$, we get
 $$\| \hat f - f^0 \|_n^2 +   \lambda \| \hat \beta_{S_2 \cup S_3} -(b^S )_{S_2}  \|_1 
 \le 2II, $$
 and hence
  $$\| \hat f - f^0 \|_n^2 +   \lambda \| \hat \beta -b^S   \|_1 
 \le {7 \over 3} II $$
 $$= {28 \over 3}   \lambda \| (b^S)_{S_2} \|_1 + {7 \over 6}
\biggl ( {\lambda_0 \over  \lambda^{\alpha} }\biggr )^{2 \over 1- \alpha } 
 + 7  \| {\rm f}_S - f^0 \|_n^2. $$
 \hfill $\sqcup \mkern -12mu \sqcap$
 
  {\bf Proof of Lemma \ref{eigenvalue-entropy}.}
  Let $\| \beta \|_1 = 1$. Then 
  $\| \beta \|_2  \le 1  $, and hence $\| E^T \beta \|_2 \le 1$.
  For $N \le V^{-1} (\delta)$ it holds that
  $\omega_{N+1} \le \delta $ and hence
  $$ \sum_{j=N+1}^p \omega_j^2 ( E^T \beta)_j^2 \le
  \omega_{N+1}^2  \sum_{j=N+1}^p ( E^T \beta)_j^2 \le \delta^2.  $$
  We now note that $\| \beta \|_1 = 1$ implies
  $\| f_{\beta } \|_n \le 1$
  and hence 
  $$\sum_{j=1}^N \omega_j^2 ( E^T \beta)_j^2 \le 1 . $$
  Lemma 14.27 in \cite{BvdG2011} states that a ball with radius $1$
  in $N$-dimensional Euclidean space can be covered
  by $(3/\delta)^N $ balls with radius $\delta$
  (see also Problem 2.1.6 in
 \cite{vanderVaart:96}).
  \hfill $\sqcup \mkern -12mu \sqcap$
  
%  {\bf Proof of Lemma \ref{compa.lemma}.} 
%Let for $k=1 , \ldots , N$,
%$$V_k := \{ j: \|  \psi_j - \phi_k \|_n = \min_{1 \le l \le N}  \| \psi_j - \phi_l \|_n\} . $$
%Since $S \not= \{ j_1 , \ldots , j_N \}$
%there exists a $\psi_j \notin \{ \phi_k\}_{k=1}^N $. We must have
%$\psi_j \in V_k$ for some $k$. Hence, $\| \psi_j - \psi_k \|_n \le u^2$. 
%Choose $\beta_{j,S} = {\rm l} \{ j = j_1\} $  with $\psi_{j_1} = \phi_k $, and
%$\beta_{j, S^c}= {\rm l} \{ j = j_2 \} $ with $j_2 \in V_k $.
%\hfill $\sqcup \mkern -12mu \sqcap$

%{\bf Proof of Lemma \ref{measurecorr.lemma}.} Let $\{ \phi_k \}_{k=1}^M  \subset \{ \psi_j \} $ be a maximal set
%such that $| \rho (\phi_k, \phi_l  )|  < 1- u^2 $. Then for all
%$\psi_j \notin \{ \phi_k \} $, we must have for some $k$
%$$ |\rho (\psi_j , \phi_k ) | \ge 1-u^2  , \ . $$
%But then either 
%$$ \| \psi_j - \phi_k \|_n^2 \le  2 u^2 $$
%or
%$$ \| \psi_j + \phi_k \|_n^2 \le  2u^2 . $$
%Hence $\{ \pm \phi_{k} \} $ is a $\sqrt 2 u$-covering set of
%$\{ \pm \psi_j \}$, with cardinality
%$$|  \{ \pm \phi_k \} | = 2M(1- u^2) +1. $$
%\hfill $\sqcup \mkern -12mu \sqcap$

\bibliographystyle{plainnat}
\bibliography{reference}

\end{document}